\documentclass[superscriptaddress, aps, prb, preprint]{revtex4-1}
 
\usepackage{booktabs}
\usepackage{graphicx}
\usepackage{pdfsync}
\usepackage{natbib}
\usepackage{amsmath}%
\usepackage{amsfonts}%
\usepackage{amssymb}%
\usepackage{array}
\usepackage[caption=off]{subfig}
 
\usepackage{checkend}	
\usepackage{xspace}
\usepackage{color}
\definecolor{greytext}{gray}{0.5}

\usepackage[unicode=true,
  linktocpage,
  linkbordercolor={0.5 0.5 1},
  citebordercolor={0.5 1 0.5},
  linkcolor=blue]{hyperref}

\newcommand{\YBCO}[1]{YBa\ensuremath{_2}Cu\ensuremath{_3}O\ensuremath{_{6{.#1}}}\xspace}

\newcommand{\caxis}{\ensuremath{\hat{c}}-axis\xspace}

\newcommand{\Tc}{\ensuremath{T_{c}}\xspace}
\newcommand{\highTc}{high-\ensuremath{T_c}\xspace}

\newcommand{\pcg}{PuCoGa\ensuremath{_5}\xspace}%
\newcommand{\cci}{CeCoIn\ensuremath{_5}\xspace}

\newcommand{\eg}{e.g.,\ }	
\newcommand{\ie}{i.e.,\ }	

\begin{document}

\title{Avoided Valence Transition in a Plutonium Superconductor}

\let\clearpage\relax
\author{B.~J.~Ramshaw}
\affiliation{Los Alamos National Labs, Los Alamos, NM, United States of America}
\author{A.~Shekhter}
\affiliation{Los Alamos National Labs, Los Alamos, NM, United States of America}
\author{R.~D.~McDonald}
\affiliation{Los Alamos National Labs, Los Alamos, NM, United States of America}
\author{J.~B.~Betts}
\affiliation{Los Alamos National Labs, Los Alamos, NM, United States of America}
\author{J.~N.~Mitchell}
\affiliation{Los Alamos National Labs, Los Alamos, NM, United States of America}
\author{P.~H.~Tobash}
\affiliation{Los Alamos National Labs, Los Alamos, NM, United States of America}
\author{C.~H.~Mielke}
\affiliation{Los Alamos National Labs, Los Alamos, NM, United States of America}
\author{E.~D.~Bauer}
\affiliation{Los Alamos National Labs, Los Alamos, NM, United States of America}
\author{A.~Migliori}
\affiliation{Los Alamos National Labs, Los Alamos, NM, United States of America}

\date{Sept 13, 2014}

\maketitle

\newcommand*{\citen}{}
\DeclareRobustCommand*{\citen}[1]{%
  \begingroup
    \romannumeral-`\x 
    \setcitestyle{numbers}%
    \cite{#1}%
  \endgroup
}

\textbf{Some of the most remarkable phenomena---and greatest theoretical challenges---in condensed matter physics arise when $d$ or $f$ electrons are neither fully localized around their host nuclei, nor fully itinerant. This localized/itinerant ``duality'' underlies the correlated electronic states of the high-\Tc cuprate superconductors and the heavy-fermion intermetallics, and is nowhere more apparent than in the $5f$ valence electrons of plutonium. Here we report the full set of symmetry-resolved elastic moduli of \pcg---the highest \Tc superconductor of the heavy fermions ($T_c$=18.5 K)---and find that the bulk modulus softens anomalously over a wide range in temperature above \Tc. Because the bulk modulus is known to couple strongly to the valence state, we propose that plutonium valence fluctuations drive this elastic softening. This elastic softening is observed to disappear when the superconducting gap opens at \Tc, suggesting that plutonium valence fluctuations have a strong footprint on the Fermi surface, and that \pcg avoids a valence-transition by entering the superconducting state. These measurements provide direct evidence of a valence instability in a plutonium compound, and suggest that the unusually high-\Tc in this system is driven by valence fluctuations. }

\pcg enters a superconducting state below $T_c =18.5$ K\cite{Sarrao:2002}---an order of magnitude higher than all Ce- or U-based superconductors. This raises the question of what makes plutonium, rather than other lanthanides and actinides, especially favourable for superconductivity. In general, the valence $f$-electrons in many lanthanide and actinide metals and compounds are nearly degenerate with the conduction band, supporting two or more nearly degenerate valence configurations\cite{Lawrence:1981}.  In some cases this valence degeneracy becomes unstable, leading to valence fluctuations and ultimately a transition to a different valence state as a function of temperature, pressure, and/or doping\cite{Kindler:1994}. X-ray and photoemission spectroscopy\cite{Joyce:2003,Booth:2014}, neutron form factor measurements\cite{Hiess:2008}, and theoretical calculations\cite{Pezzoli:2011} all indicate that \pcg is in an intermediate valence state, with the $5f^6$ (Pu$^{2+}$), $5f^5$ (Pu$^{3+}$), and $5f^4$ (Pu$^{4+}$) orbitals all residing near the Fermi level and all partially occupied.  Because of  this proximity to the conduction band, plutonium's 5$f$ electrons cannot be understood as either as fully localized nor fully itinerant\cite{Joyce:2003}.  This duality of the 5$f$ electrons is common to many plutonium compounds, and to elemental plutonium itself\cite{Albers:2001,Shim:2007}. In contrast, the analogous CeMIn$_5$ (M=Co, Rh, Ir) series of superconductors exhibits localized cerium $4f$ electrons \cite{Pagliuso:2002} and resides close to an antiferromagnetic quantum critical point\cite{Thompson:2007}, where associated antiferromagnetic spin fluctuations are believed to mediate unconventional superconductivity. \pcg, on the other hand, has quenched local moments below the nominal Kondo temperature of $T_K \approx 250(30)$~K \cite{Bauer:2004}, and a small and weakly temperature-dependent magnetic susceptibility (see footnote\footnote{The discovery paper for \pcg reported strong Curie-Weiss susceptibility, consistent with a local moment. However, further susceptibility measurements (including polarized neutron scattering) have not reproduced this \cite{Hiess:2008}: the original indication of magnetism may have come from an impurity phase.}). Because there appear to be no local moments in \pcg, and no evidence for proximity to a magnetically ordered state\cite{Bauer:2012}, antiferromagnetic spin fluctuations are unlikely to be the driver of the anomalously high \Tc in \pcg, suggesting a different origin for its high-\Tc superconductivity. Here, we report a softening of the elastic moduli over a large temperature range above \Tc, reflecting the presence of valence fluctuations in \pcg.

\begin{figure}%
\begin{minipage}[]{.44\columnwidth}
\subfloat{
\includegraphics[width=\columnwidth, clip=true,trim = 0mm 0mm 0mm 0mm]{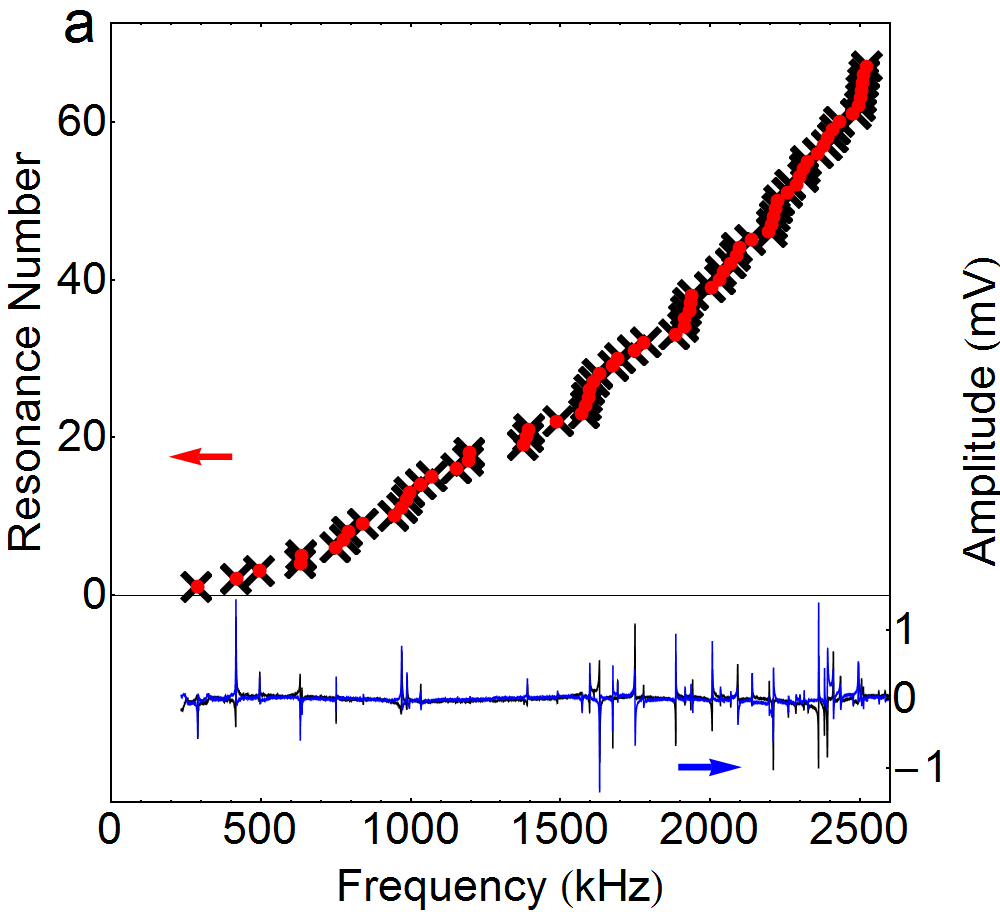}
\label{fig:1a}
}
\end{minipage}
\hspace{.02\textwidth}
\begin{minipage}[]{.50\columnwidth}
\subfloat{
\includegraphics[width=\columnwidth, clip=true,trim = 0mm -15mm 0mm 0mm]{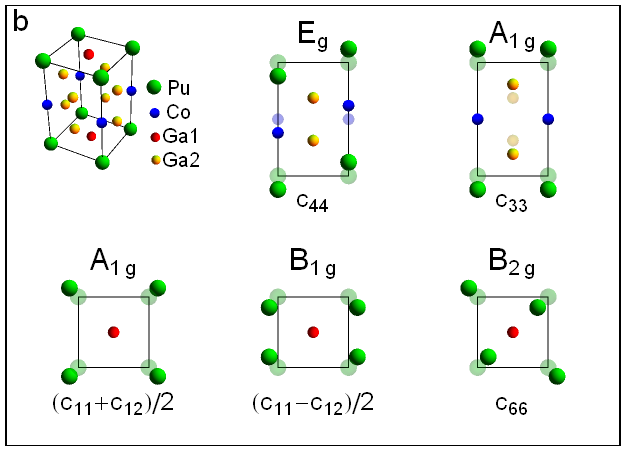}
\label{fig:1b}
}
\end{minipage}
\caption{\textbf{Vibrational spectrum of PuCoGa$_5$ at room temperature.} \textbf{a,}Transmitted ultrasonic signal as a function of frequency (real and imaginary components in blue and black: right vertical axis), showing the first 65 resonances at room temperature. The position of each resonance is indicated by a red dot (left vertical axis): the elastic moduli are calculated precisely by fitting these resonance positions---a highly overdetermined problem with six moduli and 65 resonances. The calculated resonance positions (black crosses) not only have a small residual error, but also reproduce the correct structure of the data. \textbf{b,} The \pcg unit cell, and the five irreducible representations of strain allowed by tetragonal symmetry and their associated elastic modulus. The sixth modulus $c_{13}$ is the coupling coefficient between the two $A_{1g}$ strains. These six moduli at $T=295~\mathrm{K}$ are measured to be $\left(c_{11} +c_{12} \right)/{2} = 119.8$,  $c_{13} = 64.5,$ $c_{33} = 176.1,$  $\left(c_{11} - c_{12} \right)/{2} =73.1$, $c_{44}=60.8$, and $c_{66} = 54.2$ (all values in GPa).}
\label{fig:1}
\end{figure}

\begin{figure}%
\subfloat{
\includegraphics[width=0.42\textwidth, clip=true, trim =10mm 10mm 20mm 30mm]{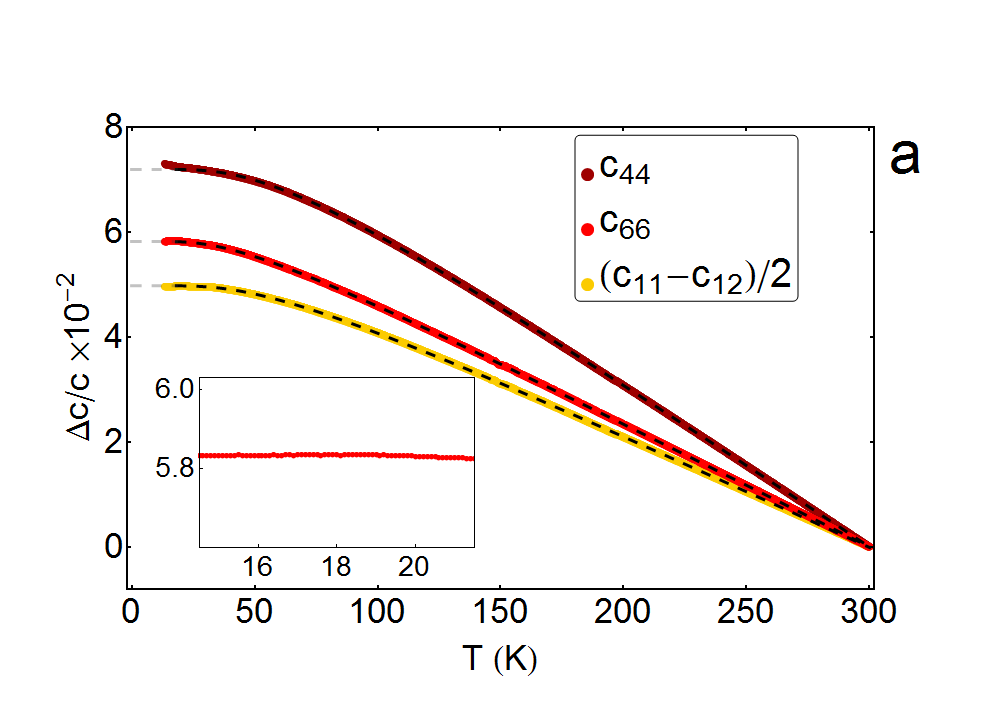}
\label{fig:2a}
}
\subfloat{
 \includegraphics[width=0.42\textwidth,clip=true, trim = 10mm 10mm 20mm 30mm]{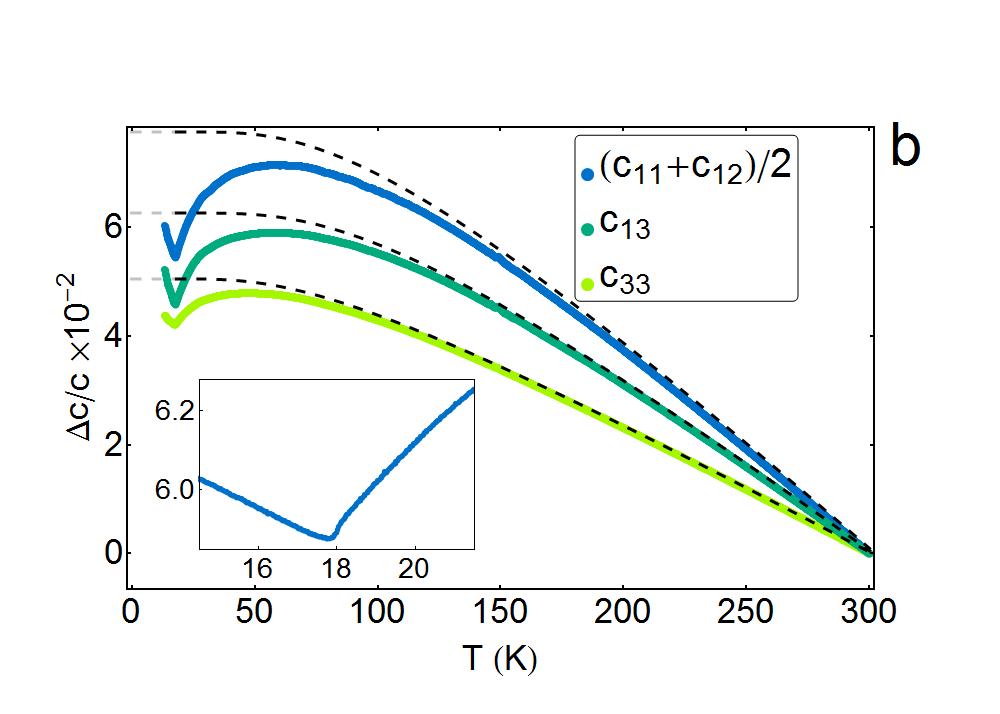}
\label{fig:2b}
}
\linebreak
\subfloat{
\includegraphics[width=0.42\textwidth, clip=true,trim = 10mm 10mm 20mm 30mm]{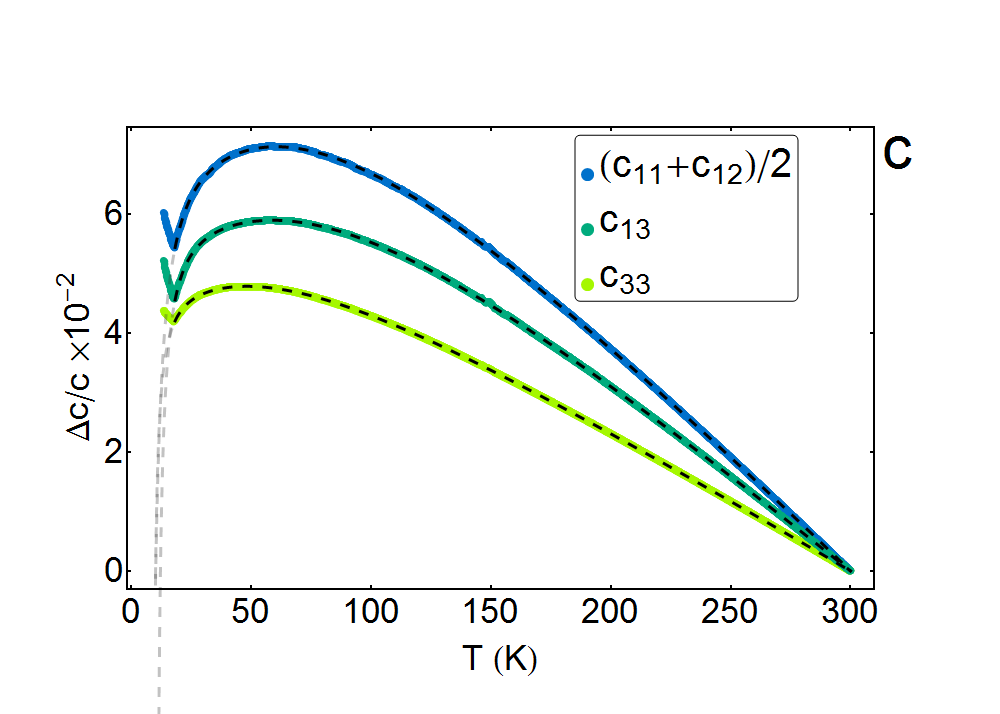}
\label{fig:2c}
}
\subfloat{
\includegraphics[width=0.42\textwidth,clip=true, trim = 10mm 10mm 20mm 30mm]{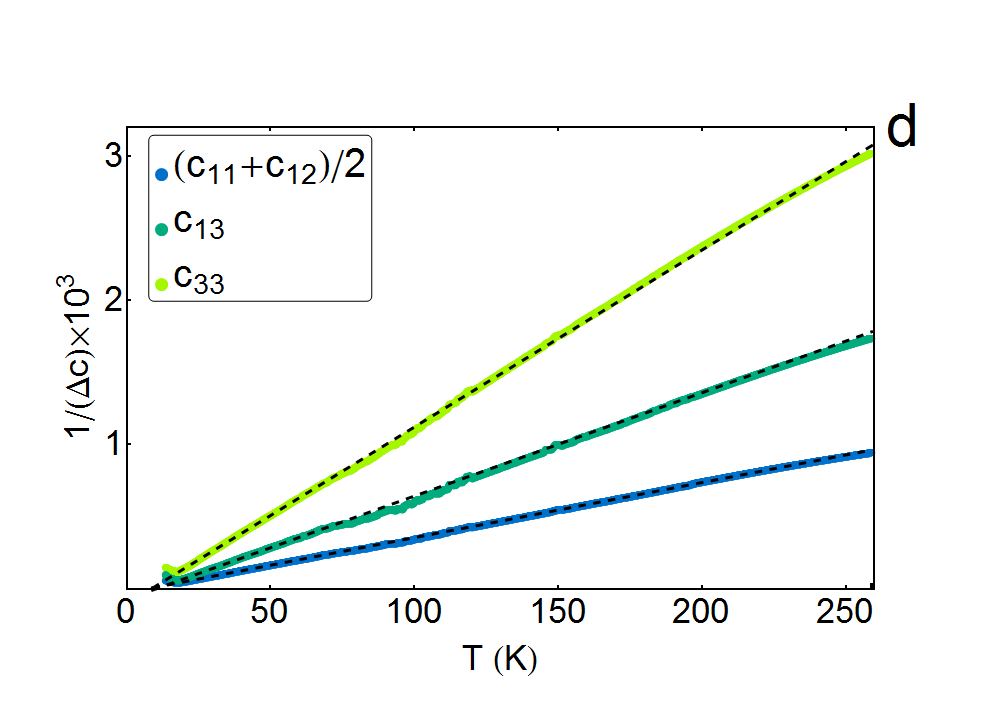}
\label{fig:2d}
}
\caption{\textbf{Temperature dependence of the elastic moduli of \pcg.} \textbf{a,} Shear moduli, normalized to the room temperature values. The dashed black line represents a three parameter fit to the standard temperature dependence from an anharmonic lattice, $c\left(T\right) = a - s/\left(e^{T_D/T} -1\right)$ (Ref. \citen{Migliori:2008}).  \textbf{b,} Scalar moduli show softening across a broad temperature range and truncating at \Tc. The dashed black line represents the anharmonic background, from which the data deviate strongly at low temperature. \textbf{c,} The scalar moduli from panel \textbf{b} with a fit to both the anharmonic background plus a $-a/\left(T_v-T\right)$ contribution, where $T_v = 9\pm 1.0$ K. \textbf{d} The inverse of the difference between the scalar moduli and the anharmonic background: this residual is linear above \Tc, and the intercept is  $9\pm 1.5$ K.  }%
\label{fig:2}%
\end{figure}

\begin{figure}[h!]%
\subfloat{
\includegraphics[width=.32\columnwidth,clip=true, trim = 5mm 20mm 2mm 5mm]{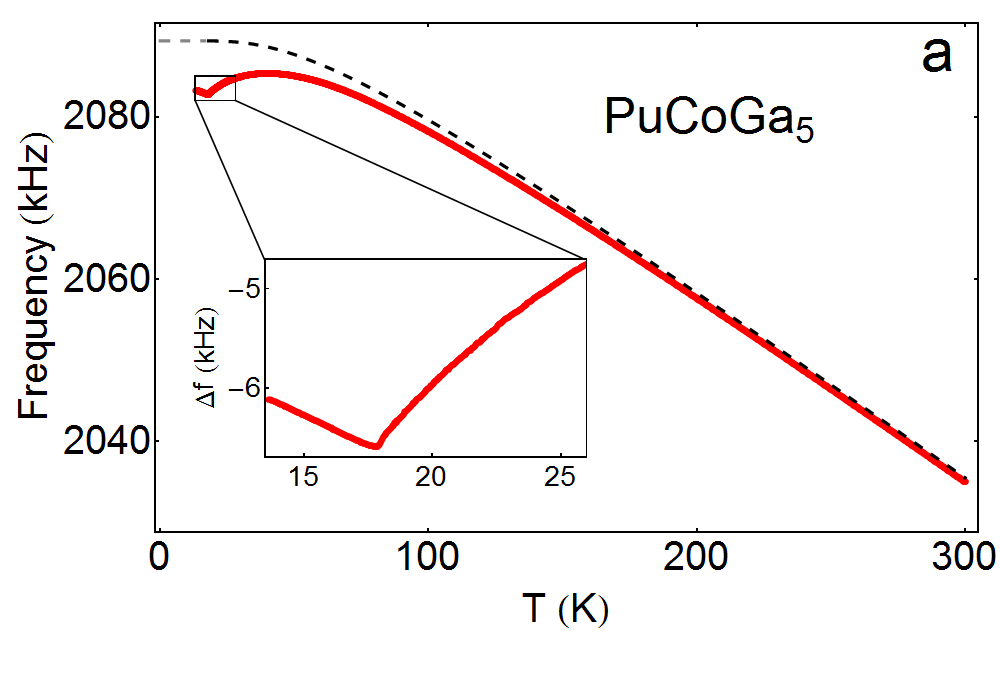}
}
\subfloat{
\includegraphics[width=.32\columnwidth, clip=true,trim =  5mm 20mm 2mm 5mm]{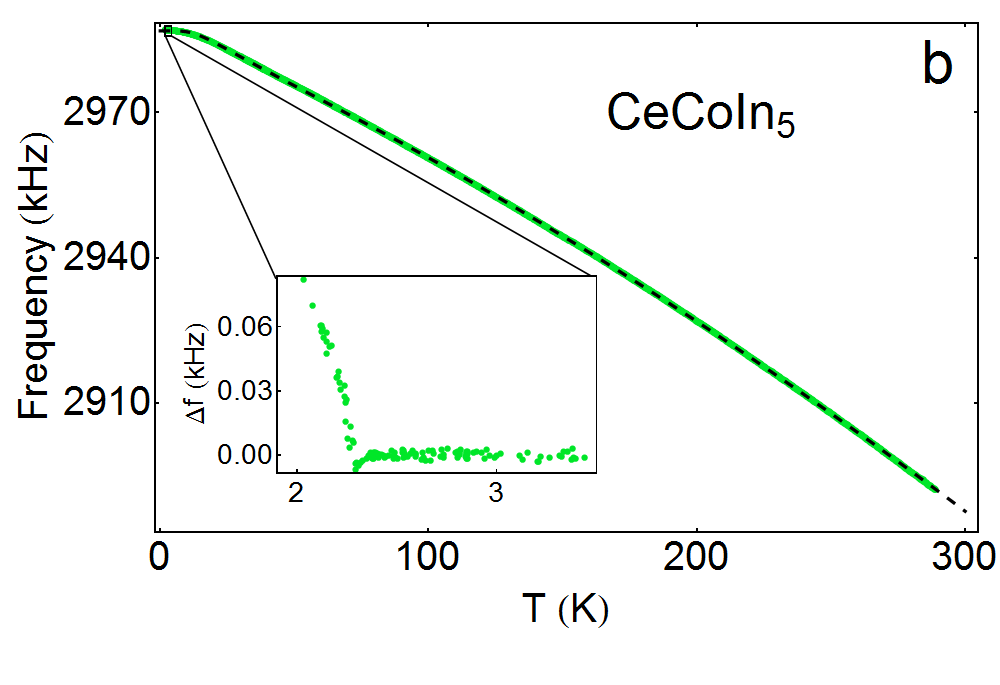}
}
\subfloat{
\includegraphics[width=.32\columnwidth,clip=true,trim = 5mm 20mm 2mm 5mm]{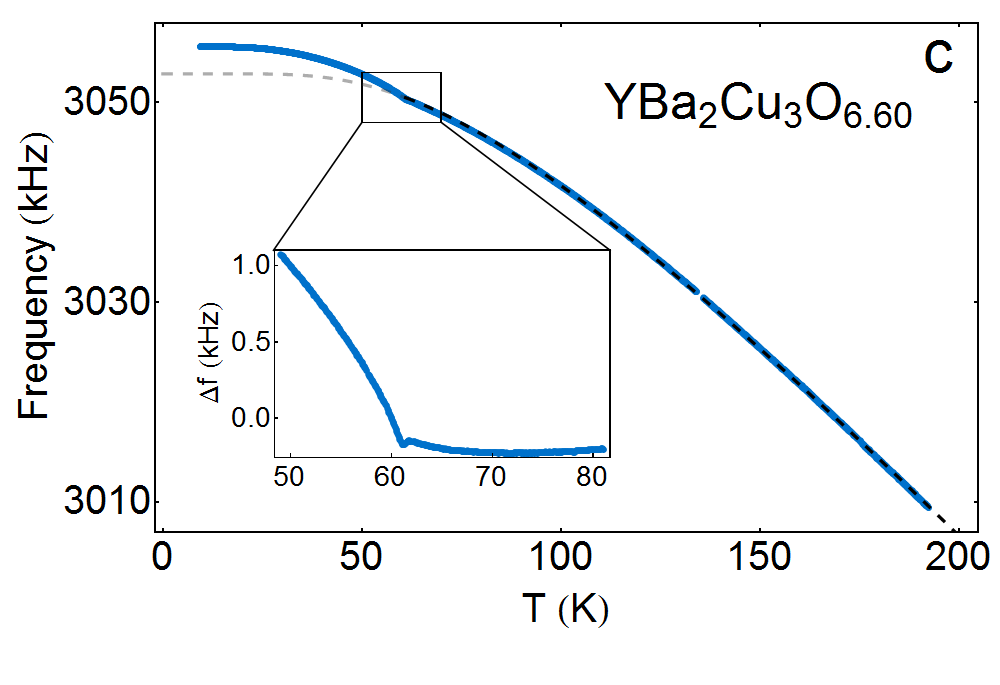}
}
\caption{\textbf{Compressional resonances in three unconventional superconductors.} Resonance modes dominated by the scalar moduli are shown for \pcg (\textbf{a}), \cci (\textbf{b}), and \YBCO{60} (\textbf{c}). While all three materials show a discontinuity at \Tc, and all stiffen by about $\Delta f / f \approx 5\times 10^{-5}/$K immediately below \Tc, only \pcg shows the dramatic softening above \Tc.}%
\label{fig:3}%
\end{figure}

\begin{figure}[h!]%
\begin{minipage}[]{.41\columnwidth}
\centering
\subfloat{
\includegraphics[width=.95\columnwidth, clip=true, trim = 0mm 40mm 20mm 40mm]{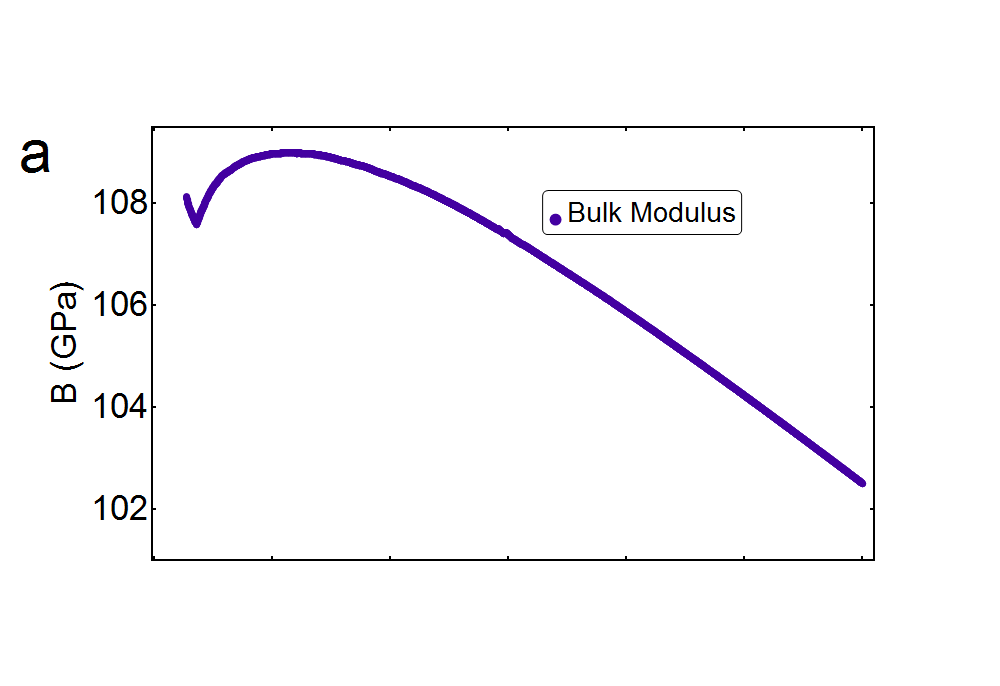}
\label{fig:4a}
}
\vspace{-3mm}
\subfloat{
\includegraphics[width= .95\columnwidth,clip=true, trim = 0mm 40mm 20mm 40mm]{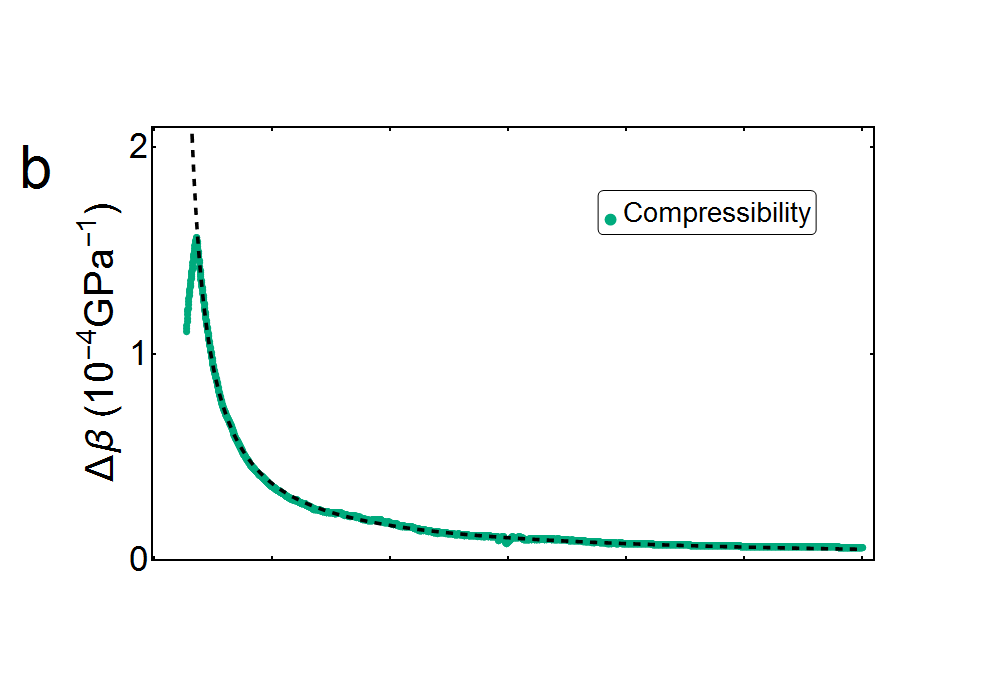}
\label{fig:4b}
}
\vspace{-3mm}
\subfloat{
\includegraphics[width= .95\columnwidth,clip=true, trim = 0mm 40mm 20mm 40mm]{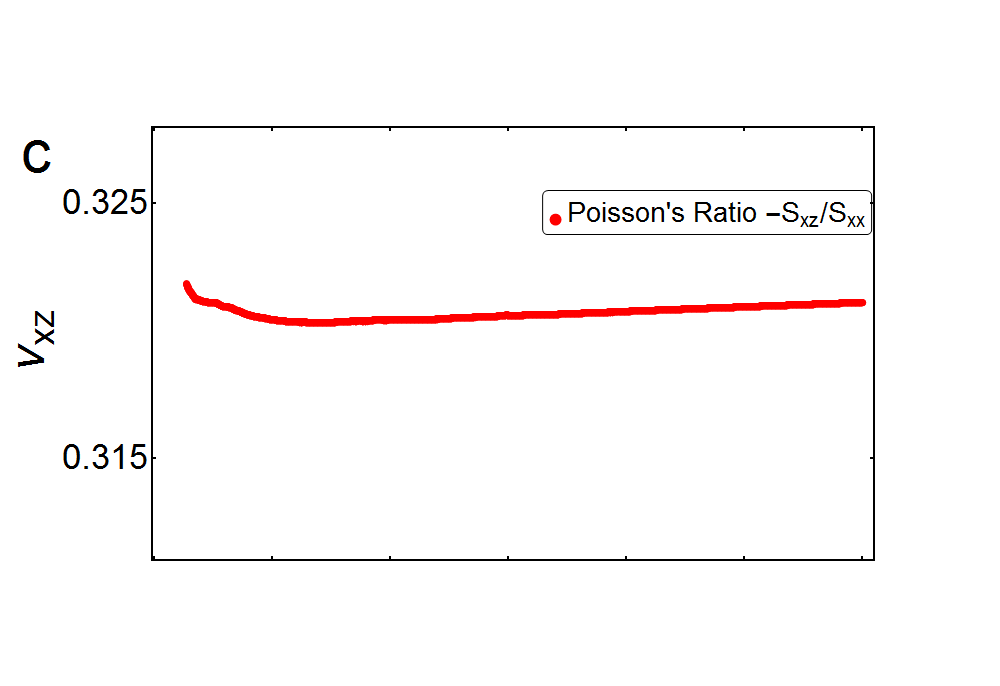}
\label{fig:4c}
}
\vspace{-3mm}
\subfloat{
\includegraphics[width= .95\columnwidth,clip=true, trim = 0mm 10mm 20mm 40mm]{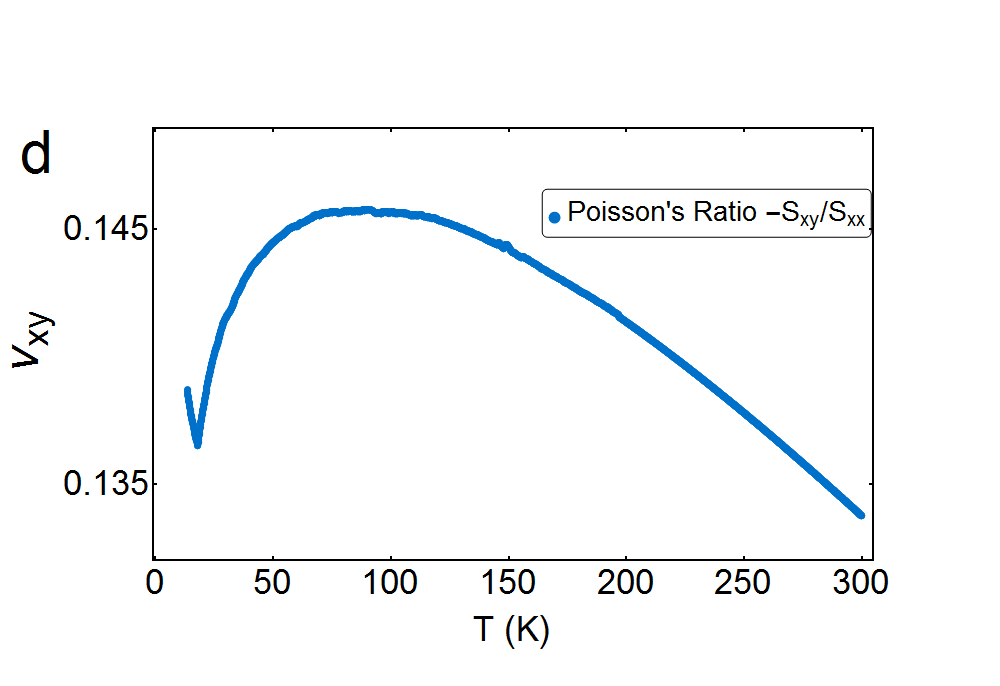}
\label{fig:4d}
}
\end{minipage}
\begin{minipage}[]{.57\columnwidth}
\centering
\subfloat{
\includegraphics[width=.5\columnwidth, clip=true, trim = 0mm 25mm 0mm 10mm]{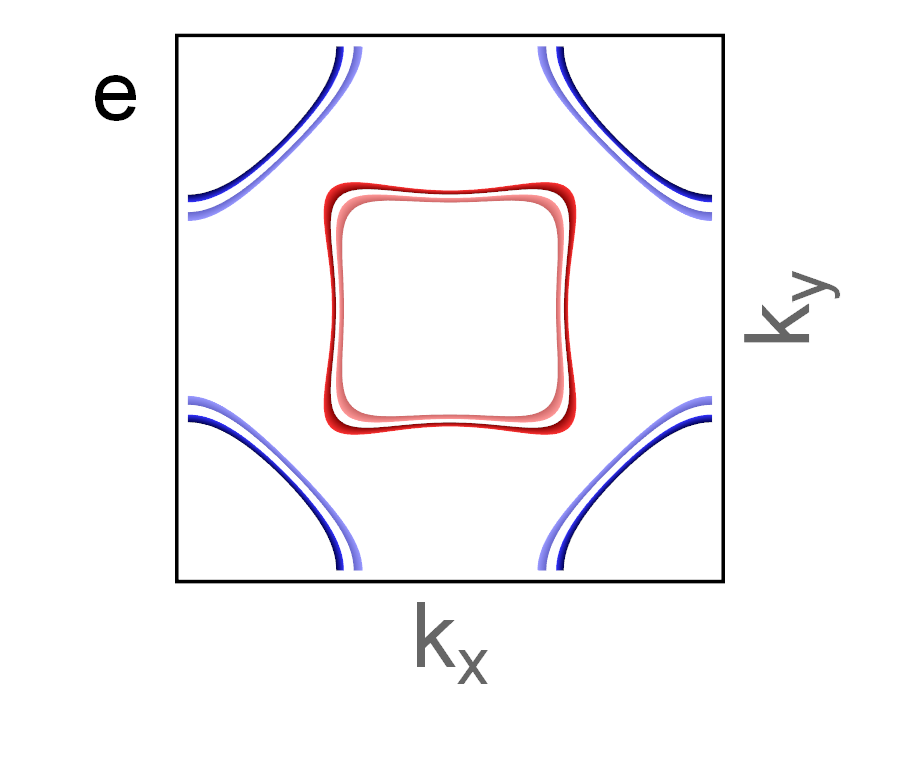}
\label{fig:4e}
}
\subfloat{
\includegraphics[width= .5\columnwidth, clip=true, trim = 0mm 25mm 0mm 10mm]{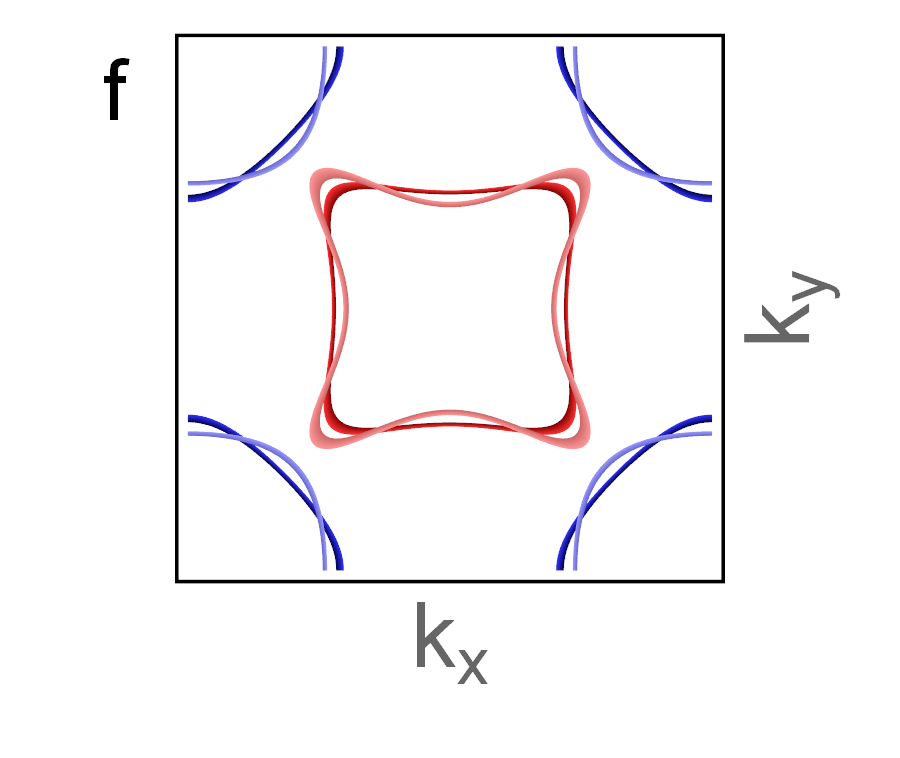}
\label{fig:4f}
}
\end{minipage}

\caption{
\textbf{Valence fluctuations in \pcg.} \textbf{a,b,c,d,} the bulk modulus, anomalous compressibility (inverse of the bulk modulus), out-of-plane Poisson's ratio ($\nu_{xz} \equiv -S_{xz}/S_{xx}$, where $S_{ij}$ are elastic compliances from the inverted modulus tensor), and in-plane Poisson's ratio ($\nu_{xy} \equiv -S_{xy}/S_{xx}$), respectively. The Poisson's ratio $\nu_{xz}$ has a magnitude typical of most metals\cite{Greaves:2011} and is only weakly temperature dependent. The ratio $\nu_{xy}$, on the other hand, is strongly temperature dependent and anomalously small, reminiscent of other mixed-valence systems\cite{Thalmeier:1991}. The anomalous contribution to the compressibility is is shown in \textbf{b}, along with a fit to  $-a/\left(T_v-T\right)$, where $T_v=9\pm1$ K. \textbf{e,f,} Schematic Fermi surface of \pcg (after \citet{Zhu:2012} and \citet{Maehira:2006}), with hole-pockets in red and electron-pockets in blue. $A_{1g}$ symmetry Fermi surface distortions are shown in lighter shades, with fluctuations of the total carrier density in \textbf{e}, and a volume-preserving distortion due to hybridization fluctuations in \textbf{f}. 
} %
\label{fig:4}%
\end{figure}

Elastic moduli measurements are a powerful tool for revealing valence instabilities and transitions\cite{Lawrence:1981,Luthi:1985}. Recent advances\cite{Shekhter:2013} in resonant ultrasound spectroscopy (RUS), further extended in this work (see section II of the S.I.), have allowed us to resolve all the elastic moduli of \pcg to low temperature in a single temperature sweep. This provides a unique opportunity to explore the unusual valence of plutonium with a thermodynamic probe that is sensitive to symmetry. \autoref{fig:1a} shows the first 65 resonance modes---the lowest-energy vibrational excitations---of a $2.208 \times 2.240 \times 0.641$ mm single crystal of \pcg (see Section I of the S.I. for experimental details; see footnote for sample ageing information\footnote{The \Tc of freshly-grown \pcg is 18.5 K. As the sample self-irradiates due to the decay of plutonium, this \Tc decreases at a rate of 0.25 K per month, resulting in our sample having a \Tc of 18.1 K. We checked the elastic moduli over the period of one month and observed no qualitative changes in the moduli induced by radiation damage, other than the decrease of \Tc.}). Each resonance frequency is uniquely determined by crystal geometry, density, and six elastic moduli---a consequence of the five irreducible strains in this tetragonal system (see \autoref{fig:1b}); conversely, the measured resonance frequencies uniquely determine the six elastic moduli. By fitting the 65 resonance frequencies as a function of temperature, we extract the temperature dependencies of the elastic moduli from room temperature to below the superconducting transition, as shown in \autoref{fig:1} (an example of the fit at $T=295~\mathrm{K}$ is shown in \autoref{fig:1a}; see Section II of the S.I. for details of the data analysis). There are three shear moduli associated with volume-preserving strains (transforming as $B_{1g}$, $B_{2g}$, and $E_{g}$ irreducible representations), shown in \autoref{fig:2a}, and three ``bulk'' moduli associated with volume-changing strains (all transforming as the $A_{1g}$ representation, which we will refer to as ``scalar'' because they preserve the lattice symmetry), shown in \autoref{fig:2b}. The scalar moduli behave very differently from the shear moduli: the shear moduli show no anomalies over the entire temperature range (including through \Tc) and their temperature dependence is described by the standard Einstein-oscillator model for an anharmonic lattice\cite{Migliori:2008} (linear at high temperature, constant low temperature, see \autoref{fig:2a}). In contrast, the scalar moduli fall below the anharmonic background well above \Tc, as shown in \autoref{fig:2d} (note that the bulk modulus (\autoref{fig:4a}) is a particular combination of scalar moduli for hydrostatic strain). This softening in all three scalar moduli follows $\sim 1/\left(T - T_v\right)$ behaviour, where $T_v \approx 9~\mathrm{K}$, as seen in \autoref{fig:2c}. This softening is truncated when superconductivity sets in at $\Tc=18.1$ K, before this nominal valence transition at $T_v \approx 9~\mathrm{K}$. This softening is not observed in either the related compound \cci, where the Ce $4f$ electrons are localized, or in the \highTc superconductor \YBCO{58} (see \autoref{fig:3}). However, below \Tc, the elastic moduli of \pcg behave similarly to these other unconventional superconductors, suggesting that this anomalous behaviour is confined to temperatures above \Tc (\autoref{fig:3}).

\autoref{fig:2d} shows that the deviation from the anharmonic background extends over a broad temperature range for all three scalar moduli. The $\sim 1/\left(T - T_v\right)$ softening of the scalar moduli reveals the existence of a fluctuating order parameter $\eta$ that couples linearly to scalar strain $\epsilon_{A_{1g}}$ (\ie $\Delta F \propto \epsilon_{A_{1g}}\!\cdot\!\eta$). This is analogous to the $\chi \propto 1/\left(T - T_c\right)$ Curie-Weiss susceptibility of a ferromagnet, where an applied magnetic field $\vec{H}$ couples linearly to the magnetic order parameter $\vec{M}$ in the free energy, \ie $\Delta F = -\vec{M}\!\cdot\!\vec{H}$.  Linear coupling demands that the order parameter $\eta$ is non-magnetic and scalar ($A_{1g}$), \ie that it has the same symmetry as the strain (see Section III of the S.I.)\cite{Rehwald:1973,Callen:1963}. As in other mixed valence systems that show scalar elastic softening, the symmetry of this order parameter $\eta$ suggests a valence instability\cite{Lawrence:1981,Luthi:1985,Kindler:1994}. Valence fluctuations lead to an anomalous temperature dependence of the scalar elastic moduli  because changes in the relative occupation of the plutonium $5f$ valence can change the unit cell volume\cite{Anderson:1974, Varma:1975} (or derivatives of the free energy with respect to volume). This is more easily visualized as a divergence in the compressibility (inverse of the bulk modulus), which we show with a fit in \autoref{fig:4b}.  It is important to note that the (complex) superconducting order parameter, $\Psi \equiv \left|\Psi\right|e^{i \phi}$, cannot be responsible for the scalar softening: $\Psi$ cannot couple linearly to any strain, but instead couples at least quadratically (scalar strain couples to superfluid density, \ie $\Delta F \propto \epsilon_{A_{1g}}\!\cdot\!n_s$,  where $n_s \equiv \left|\Psi\right|^2$). Further, the $\sim 1/\left(T - T_v\right)$ softening in \pcg is qualitatively different from the softening due to the screening of local moments in integral-valent systems. In these systems, a drop in elastic moduli is observed at the Kondo temperature, followed by saturation at lower temperature \cite{Thalmeier:1991}. We do not observe any sudden onset of softening at the Kondo temperature in \pcg ($T_K \approx 250$ \cite{Bauer:2004}), and the temperature dependence we do observe is qualitatively different from what is observed in other Kondo systems  (\eg CeCu$_6$, CeRu$_2$Si$_2$ \cite{Thalmeier:1991}). Thus we attribute the softening to valence fluctuations, similar to YbInCu$_4$, which has a valence transition at $T=65$ K, and which also shows $\sim 1/\left(T - T_v\right)$ elastic softening over a broad temperature range\cite{Kindler:1994}. 

Next we consider the behaviour of the elastic moduli across the superconducting transition. A sharp drop in the scalar moduli at \Tc (see inset of \autoref{fig:2b}) is of order $\Delta c / c \approx 3\times 10^{-4}$, within a factor of 3 of the estimate made from the specific heat jump and $\partial \Tc / \partial P$ using Erhenfest relations\cite{Sarrao:2002, Griveau:2004}, suggesting that conventional thermodynamics apply for the superconducting transition in \pcg. Upon entering the superconducting state at 18.1 K the softening in the $A_{1g}$ channel is truncated (\autoref{fig:2b}), indicating that the opening of the superconducting gap on the Fermi surface suppresses the valence fluctuations. Below the superconducting transition the elastic moduli \pcg stiffens at a rate similar to other unconventional superconductors that show no anomalous softening (\autoref{fig:3}). One can draw an analogy here with superfluid $^3$He, where the Cooper pairing is mediated by spin fluctuations, and where these fluctuations are truncated upon entering the superconducting state\cite{Anderson:1973,Wheatley:1975,Leggett:1975}.

The softening of scalar elastic moduli in mixed-valence systems is often accompanied by an anomalously small and/or strongly temperature-dependent Poisson's ratio\cite{Luthi:1985} (\eg YbInCu$_4$\cite{Kindler:1994}). In a conventional material, compression along one axis produces a dilation strain along the perpendicular axes, and the ratio of the perpendicular strains is the Poisson's ratio. In a mixed-valence system, compression can force the nearly-degenerate valence orbitals to adopt a new configuration (\eg by increasing the hybridization of the $f$-electrons with the conduction band). This results in an anomalous elastic response to uniaxial strain, and a small or even negative Poisson's ratio. \autoref{fig:4c} and \autoref{fig:4d} show the temperature dependences of the two Poisson's ratios: ($\nu_{xy}$) describes the in-plane strain; ($\nu_{xz}$) describes the out-of-plane strain. The magnitude of $\nu_{xz}$ for \pcg is typical of most metals\cite{Greaves:2011}, and is nearly temperature independent (\autoref{fig:4c}). The magnitude of $\nu_{xy}$, on the other hand, is anomalously small, and its temperature dependence mirrors that of the scalar moduli (also note that the softening in $c_{33}$ is much smaller than in $(c_{11}+c_{12})/2$, see \autoref{fig:2b}). This anomalous anisotropic behaviour of the Poisson's ratios implies an anisotropic character to the valence fluctuations in \pcg.

Valence fluctuations can manifest as ``hybridization fluctuations''\cite{Sherrington:1975}, where fluctuations between 5$f$ orbitals of different in-plane directional character (\eg $f_{xyz}$ vs. $f_{z\left(x^2-y^2\right)}$) result in fluctuations of the hybridization with neighbouring Ga atoms (\autoref{fig:1b}). This idea is further supported by recent resonant X-ray emission spectroscopy (RXES) measurements on \pcg and its sister compound PuCoIn$_5$\cite{Booth:2014}. These measurements delineate an intermediate valence state for \pcg, where a dominant $5f^5$ configuration (Pu$^{3+}$ valence, 62\% weight) is degenerate with $5f^4$ (Pu$^{4+}$, 29\%) and $5f^6$ (Pu$^{2+}$, 9\%), resulting in an average valence $z \approx 3.2$. In PuCoIn$_5$, which has a 9\% longer \caxis and 8\% longer $\hat{a}$- and $\hat{b}$-axes than \pcg, the configurational weight among the 5$f$ orbitals is distributed differently: 77\% of $5f^5$, 21\% of $5f^4$, and 2\% of $5f^6$, with the same average valence of $z \approx 3.2$. Thus, if \pcg, with its smaller unit cell, is analogous to PuCoIn$_5$ under strain, these measurements suggest that the average valence ($z \approx 3.2$) remains constant under scalar strain while the distribution among the $5f$ orbitals changes. \pcg is composed of planes of Pu surrounded by octahedrally-coordinated Ga, with each Pu-Ga plane separated by  a plane of Co, and this two-dimensionality is reflected in the band-structure\cite{Maehira:2006}. Thus, fluctuations between different 5$f$ states that preserve total valence have the largest effect on the in-plane hybridization, providing a natural explanation for why $\nu_{xy}$ has a strong valence-fluctuation signature, while $\nu_{xz}$ does not. These hybridization fluctuations can be visualized as distortions of the Fermi surface shape \autoref{fig:4f} (as opposed to fluctuations of the total valence and itinerant electron number, as in \autoref{fig:4e}). This footprint on the Fermi surface is naturally consistent with the observed truncation of valence fluctuations when the superconducting gap opens at \Tc.

Our direct observation of valence fluctuations in \pcg, and the low temperature of the avoided valence transition  ($T_v\approx9$ K), suggests the proximity of a valence quantum-critical point in the \pcg pressure-temperature phase diagram. This in turn suggests that the high superconducting transition temperature in \pcg is driven (or enhanced) by valence fluctuations, in contrast to the CeMIn$_5$ materials \cite{Thompson:2007}, in which the antiferromagnetic spin fluctuations mediate superconductivity with a much lower \Tc ($\sim$ 2 K). Our observations are consistent with previous suggestions for the possibility of valence-fluctuation mediated superconductivity in \pcg \cite{Miyake:2007,Flint:2008,Bauer:2012}. A similar scenario has been proposed for CeCu$_2$Si$_2$, where one superconducting dome forms around an antiferromagnetic quantum critical point at low pressures, and a second, higher-\Tc dome forms around a valence quantum critical point at higher pressures \cite{Yuan:2003}. Quantum critical points associated with charge, rather than spin, degrees of freedom have also been proposed to explain the high-$T_c$s in the cuprates (nematic order\cite{Kivelson:1998}, current loop order\cite{Varma:1999,Chakravarty:2001}, or more recently charge-density wave order) and in the iron pnictides (nematic order\cite{Xu:2008}). We have demonstrated that, in \pcg, the charge fluctuations are directly observable experimentally. What makes \pcg attractive for the study of charge-driven quantum criticality is that there is no proximity to magnetism, and no disorder-inducing doping required to reach the highest $T_c$s. Further exploration of this nominal valence quantum critical point is warranted, particularly in magnetic fields where superconductivity can be suppressed to reveal the valence physics in the underlying metallic state.

\section{Acknowledgments}
\label{se:Acknowledgments}
The authors would like to thank , A. Finkel'stein, I. Fisher, Z. Fisk, L.P. Gorkov, J. Lawrence, J. Smith,  and J. Thompson for helpful discussions. Work at Los Alamos National Laboratory was performed under the auspices of the U.S. DOE, OBES, Division of Materials Sciences and Engineering, and the LANL LDRD Program

\end{document}